\documentclass[12pt,preprint]{aastex}
\def\beq{\begin{equation}}
\def\enq{\end{equation}}
\def\ms{$M_\odot$}

\shorttitle{}
\shortauthors{}
\begin{document}

\title{Hard X-ray emission and $^{44}$Ti line features of Tycho Supernova Remnant}
\author{Wei Wang$^1$ and Zhuo Li$^{2,3}$}
\affil{$^1$ National Astronomical Observatories, Chinese Academy of Sciences,
                 20A Datun Road, Chaoyang District, Beijing 100012, China;
                 wangwei@bao.ac.cn \\
      $^2$ Department of Astronomy and Kavli Institute for Astronomy and Astrophysics,
        Peking University, Beijing 100871, China; zhuo.li@pku.edu.cn \\
      $^3$ Key Laboratory for the Structure and Evolution of Celestial
       Objects, Chinese Academy of Sciences, Kunming 650011, China
      }

\begin{abstract}
A deep hard X-ray survey of the INTEGRAL satellite first detected
the non-thermal emission up to 90 keV in the Tycho supernova (SN)
remnant. Its 3 -- 100 keV spectrum is fitted
with a thermal bremsstrahlung of $kT\sim 0.81\pm 0.45$ keV plus a
power-law model of $\Gamma \sim 3.01\pm 0.16$. Based on the
diffusive shock acceleration theory, this non-thermal emission,
together with radio measurements, implies
that Tycho remnant may not accelerate protons up to $>$PeV but
hundreds TeV. Only heavier nuclei may be accelerated to the cosmic
ray spectral ``knee". In addition, we search for soft gamma-ray lines
at 67.9 and 78.4 keV coming from the decay of radioactive $^{44}$Ti
in Tycho remnant by INTEGRAL. A bump feature in the 60-90 keV energy band, potentially associated with the $^{44}$Ti line
emission, is found with a marginal significance level of $\sim$ 2.6 $\sigma$. The corresponding 3 $\sigma$
upper limit on the $^{44}$Ti line flux amounts to 1.5 $\times$ 10$^{-5}$ ph cm$^{-2}$ s$^{-1}$.
Implications on the progenitor of Tycho SN, considered to be the prototype of type Ia SN, are discussed.

\end{abstract}


\section{Introduction}

Tycho supernova (SN 1572) is a historical supernova occurring in
early November of 1572 in the constellation Cassiopeia region. This
supernova explosion event was suggested to be the Type Ia supernova,
but no convinced evidence was found (e.g., Ruiz-Lapuente 2004).
Recent spectral analysis of the light echo from the explosion
(Krause et al. 2008) has confirmed that the event was a Type Ia
supernova. Measurements of the distance of the Tycho remnant
still have large uncertainties. The kinematic distance obtained by
observing HI absorption toward the Tycho remnant gives a large
distance range of 1.7 -- 5.0 kpc (Albinson et al. 1986; Schwarz et
al. 1995; Tian \& Leahy 2011). Modeling of the observed $\gamma$-ray
emission from Tycho suggests a distance greater than 3.3 kpc (V\"olk
et al. 2008). A direct distance estimate is also given by X-ray
ejecta proper-motion observations by Chandra and Suzaku which give a
distance of 4$\pm 1$ kpc (Hayato et al. 2010).

Supernova remnants (SNRs) are generally thought to be the promising
sites of producing high energy cosmic rays (CRs) up to the energies
of $> 10^{15}$ eV. The accelerated electrons and protons can produce
non-thermal emissions observed from radio to gamma-ray bands. Non-thermal emission from the Tycho SNR has been detected in radio, X-rays and gamma-rays. Radio images also show a
shell-like morphology with enhanced emission along the northeastern
edge of the remnant (Dickel et al. 1991; Stroman \& Pohl 2009). Soft
X-ray images by Chandra reveal the non-thermal X-ray emission
concentrated on the SNR rim (Hwang et al. 2002; Bamba et al. 2005;
Warren et al. 2005), which has been interpreted as the evidence of
electron acceleration. The first evidence for hard X-ray emission
from the Tycho SNR was reported by HEAO 1 (Pravdo et al. 1979) which
suggested a photon index of $\sim 2.72$ from 5 -- 25 keV. RXTE also
reported a hard X-ray continuum up to 20 keV with a photon index of
$\sim 3$ (Petre et al. 1999). Recent detection by Suzaku HXD-PIN
detector up to 28 keV implies the possible presence of accelerated
electron up to the energy at least $\sim 10$ TeV (Tamagawa et al.
2009). The Suzaku spectrum of the Tycho SNR from 13 -- 28 keV was
described by a power-law model of photon index $\Gamma\sim 2.8\pm
0.5$. In gamma-ray bands, Fermi/LAT detected the GeV emission from
the Tycho SNR with a photon index of $\Gamma\sim 2.3\pm0.3$
(Giordano et al. 2012). And the VERITAS Cherenkov telescope
succeeded in measuring its TeV emission up to 10 TeV (Acciari et al.
2011), with a photon index of $\Gamma\sim 1.95\pm 0.81$.

The accelerated electrons emit synchrotron radiation observed from
radio to X-ray bands. But the non-thermal emissions in soft X-ray
bands generally are difficult to be discriminated from thermal
components in SNRs. In hard X-ray bands ($>10$ keV), the
observations are a direct way to probe the non-thermal emission
properties, constraining the accelerating ability of SNRs. The hard
X-ray properties of the Tycho SNR have never been studied in detail due
to the poor sensitivity above 20 keV of past missions. Suzaku made a
detection of the hard X-ray emission up to 28 keV (Tamagawa et al.
2008), but we do not know its spectral characteristics at higher
energies. The instruments onboard INTEGRAL have a better sensitivity
above 20 keV up to several hundred keV and would provide new
information on the non-thermal emission properties of the supernova
remnant.

In addition, hard X-ray studies on SNRs could search for the hard
X-ray lines from radioactive $^{44}$Ti at $\sim 68$ and 78 keV.
$^{44}$Ti is a short-lived radioactive isotope with a mean life of
85 years (Admad et al. 2006). In theories, the most plausible cosmic
environment for production of $^{44}$Ti is the $\alpha$-rich
freeze-out from high-temperature burning near the nuclear
statistical equilibrium (Woosley et al. 1973; Timmes et al. 1996).
This required high values for the entropy can be found in
supernovae. It is generally believed that core-collapse supernovae
dominate the production of radioactive $^{44}$Ti (Timmes et al.
1996). The decays of $^{44}$Ti can emit four lines at energies of
4.1, 67.9, 78.4 and 1157 keV, where the line flux at 4.1 keV is
about $20\%$ of the other lines, and the flux of 67.9 keV is about
$93\%$ the flux at 78.4 keV. Previously, the 67.9, 78.4 and 1157 keV lines
from Cas A were reported (Iyudin et al. 1994, 1999; Vink et al. 2001; Renaud et al. 2006a, 2006b).
Two hard X-ray lines at 67.9, 78.4 keV from SN 1987A were recently
discovered (Grebenev et al. 2012). These two supernovae are known to be of core-collapse origin. Tentative detections of the
1157 keV line from the Vela Junior (Iyudin et al. 1998) and the 4.1
keV line from the G1.9+0.3 (Borkowski et al. 2010) were also
reported.

Generally a Type Ia supernova like Tycho is not thought to produce a
large amount of $^{44}$Ti. However, great uncertainty still
surrounds the central driver of these Type Ia explosions, just what
kind of star explodes. Is the white dwarf that blows up near the
upper limit allowed by nature (the "Chandrasekhar mass"), or
lighter, and does the explosion result from the slow accretion of
matter from a companion star or the dynamic merger of two white
dwarfs? An important diagnostic of the models is the nucleosynthesis
that they produce. In the Chandrasekhar mass explosions, the fuel is
a mixture of carbon and oxygen and the burning produces a
distinctive set of iron-group and intermediate-mass elements.
Sub-Chandrasekhar mass models, on the other hand, have a component
of explosive helium burning which produces different set, rich in
the isotopes of calcium, titanium. Depending upon the kind of dwarfs
that come together, merging white dwarfs can give both. Therefore,
searching for the $^{44}$Ti lines in the Tycho remnant is an
important tool to probe the nature of the progenitor in this Type Ia
supernova. COMPTEL three-year search for the $^{44}$Ti signals of
galactic sources gave an upper limit (2$\sigma$) of $2\times
10^{-5}$ ph cm$^{-2}$ s$^{-1}$ (Dupraz et al. 1997; Iyudin et al.
1999). Early INTEGRAL/IBIS observations of the $^{44}$Ti hard X-ray
lines also implies an upper limit (3$\sigma$) of $1.5\times 10^{-5}$
ph cm$^{-2}$ s$^{-1}$ (Renaud et al. 2006b). In this work we will
report our detections of the hard X-ray emission up to 100 keV and searching for
$^{44}$Ti emission lines in Tycho remnant with INTEGRAL
observations.

\section{INTEGRAL Observations and data analysis}

INTEGRAL is an ESA's currently operational space-based hard
X-ray/soft gamma-ray telescope covering a wide energy range of 3 keV
-- 8 MeV (Winkler et al. 2003). In this work, we have used two main
instruments aboard INTEGRAL, the imager IBIS (Ubertini et al. 2003)
and X-ray monitors JEM-X (Lund et al. 2003). The hard X-ray data are
mainly collected with the low-energy array called IBIS-ISGRI
(INTEGRAL Soft Gamma-Ray Imager) which consists of a pixellated
128$\times 128$ CdTe solid-state detector that views the sky through
a coded aperture mask (Lebrun et al. 2003). IBIS/ISGRI has a 12'
(FWHM) angular resolution and arcmin source location accuracy in the
energy band of 15 -- 200 keV. JEM-X as the small X-ray detector
collects the lower energy photons from 3 -- 35 keV which is used to
constrain the lower hard X-ray band spectral properties of Tycho
SNR.

The Tycho SNR is frequently observed during the INTEGRAL surveys on
the Cassiopeia region. We use the available archival data from the
INTEGRAL Science Data Center (ISDC) where the Tycho SNR was within
$\sim 12$ degrees of the pointing direction of INTEGRAL/IBIS
observations. The total on-source time obtained in our analysis is
about 4.9 Ms after excluding the bad data due to solar flares and
the INTEGRAL orbital phase near the radiation belt of the Earth. The
analysis was done with the standard INTEGRAL off-line scientific
analysis (OSA, Goldwurm et al. 2003) software, ver. 10. Individual
pointings in all collected IBIS data processed with OSA 10 were
mosaicked to create the sky images for the source detection in the
energy ranges of 20 -- 60 keV and 60 -- 90 keV. The Tycho SNR was
detected by IBIS with significance levels of $11.6\sigma$ and $\sim
5.0\sigma$ in two energy ranges respectively (see middle and right
panels in Fig. 1). JEM-X imagers have a much smaller field of view
(requiring observing off-axis angle $<5^\circ$) and a relatively low
sensitivity because of small detector area, the total on-source time
for the Tycho SNR is about 460 ks. The mosaic map around Tycho
detected JEM-X is also shown in Fig. 1 (left panel). The detection
significance level is about 9.8$\sigma$ in the range of 3 -- 10 keV.

The spectral extraction processes for IBIS and JEM-X are carried out
individually. Tycho remnant is about $8'$ in diameter in the sky.
This size is smaller than the angular resolution of IBIS-ISGRI
($12'$), but larger than the angular resolution of JEM-X (3'). For
IBIS, the spectral extraction was done using the software script {\em ibis\_ science\_ analysis} up to the {\em SPE} level with the input source
catalog. Above $\sim 90$ keV, only upper limits can be given by the
ISGRI detector. For the spectral analysis on Tycho SNR using JEM-X,
we have made use of the {\em mosaic\_ spec} script to extract the spectrum
from 3 -- 35 keV by assuming a source size of $\sim 8'$. The
spectral data points are directly derived from the mosaic images of
JEM-X in four energy bands: 3 -- 6 keV, 6 -- 10 keV, 10 -- 16 keV,
and 16 -- 35 keV.

\begin{figure*}
\centering
\includegraphics[angle=0,width=18cm]{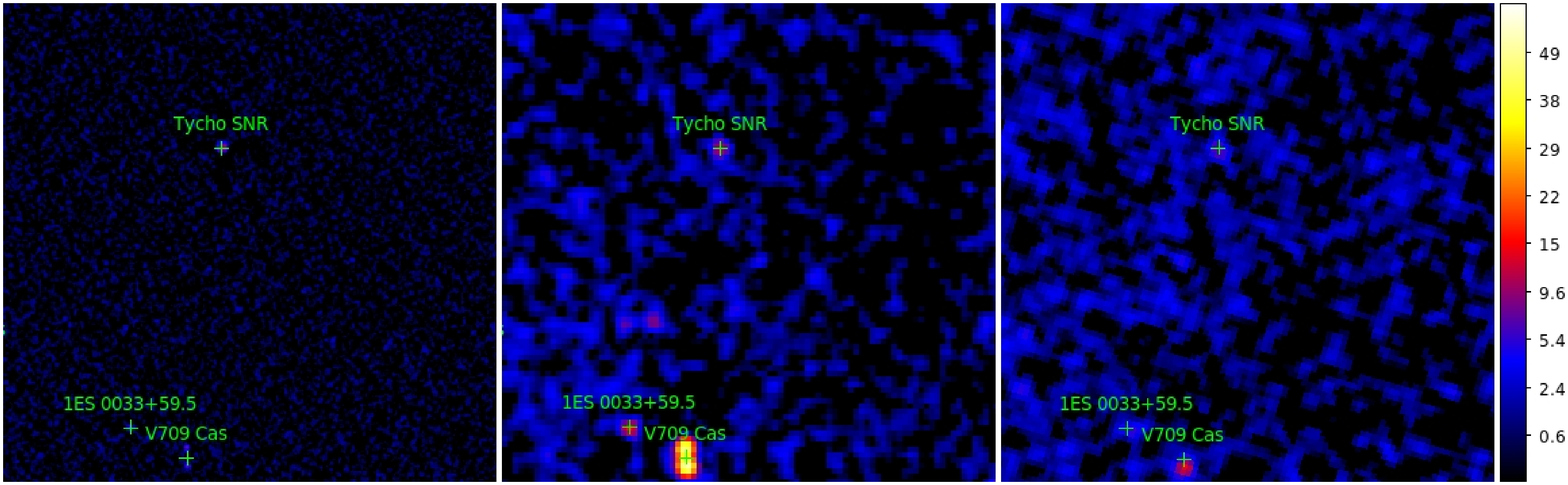}
\caption{Significance mosaic maps ($6^\circ\times 6^\circ$) around
Tycho supernova remnant in Equatorial J2000 coordinates as seen with
INTEGRAL/JEM-X (left) in the range of 3 -- 10 keV and INTEGRAL/IBIS
in the energy band of 20 -- 60 keV (middle) and 60 -- 90 keV
(right). Tycho SNR was detected by JEM-X with a significance level
of 9.8$\sigma$ and by IBIS with the significance levels of $\sim
11.6\sigma$ and $5\sigma$ in two energy bands, respectively. }
\end{figure*}

\section{Hard X-ray spectral characteristics of Tycho SNR}

\begin{figure}
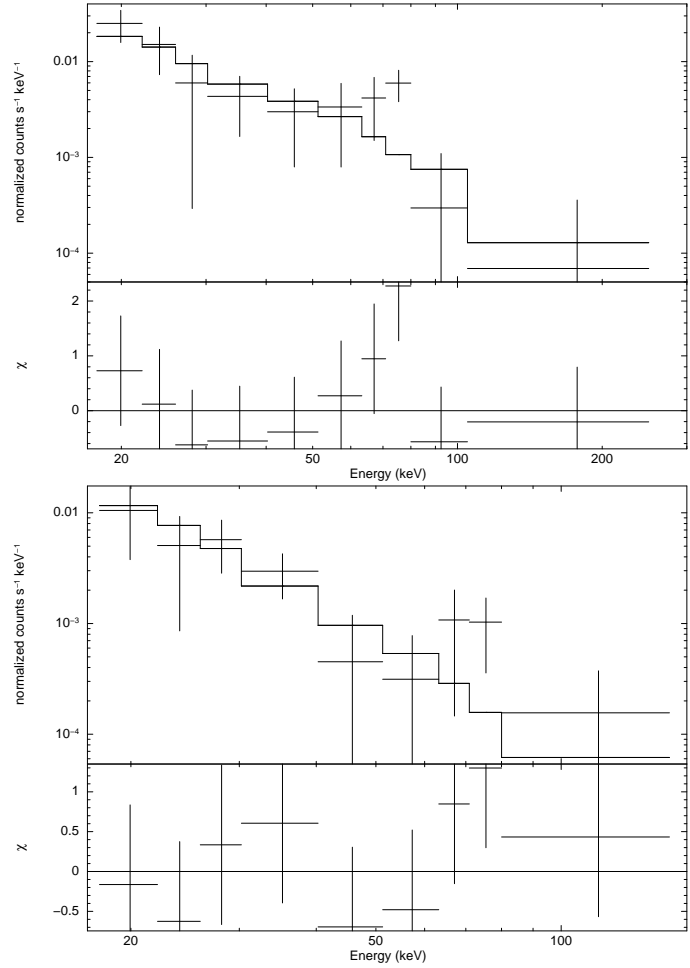

\centering
\includegraphics[angle=-90,width=9cm]{tycho_331spec.eps}
\includegraphics[angle=-90,width=9cm]{tycho_800spec.eps}
\caption{IBIS/ISGRI hard X-ray spectra of Tycho SNR for two different datasets and the associated best-fit
power-law models. {\bf Top:} data from Rev. 331 to 396 with $\Gamma$ = 2.5 $\pm$ 0.4 (reduced $\chi^2$ $
\sim$ 1.23, 8 d.o.f). {\bf Bottom:} data from Rev. 883 to 1066 with $\Gamma$ = 2.9 $\pm$ 0.5 (reduced $
\chi^2$ $\sim$ 1.13, 7 d.o.f). }
\end{figure}

We first derived the hard X-ray spectrum of the Tycho remnant
obtained by IBIS which has a very long exposure on the source. In
Fig. 2, we present the spectra of Tycho from 18 -- 150 keV in two
time intervals when IBIS carried out deep observations on the
source, one in 2005 and the other from 2010 -- 2011. Both two
spectra are fitted with a simple power-law model. There exists a
feature around 60 -- 90 keV in both spectra. These features may be
attributed to the $^{44}$Ti line signal. To probe the bump feature
near 60 -- 90 keV in details, we re-extracted the IBIS hard X-ray
spectra from 18 -- 200 keV using all the available data with smaller
energy bins from 30 -- 90 keV. The lower energy band data points can
be used to constrain the continuum better, so that the JEM-X
spectrum of Tycho is extracted for the analysis together in the
followings.

The extracted hard X-ray spectra from 3 -- 35 keV from JEM-X and 18
-- 200 keV from IBIS for the Tycho SNR are displayed together in
Fig. 3.  The spectrum from 3 -- 200 keV is initially fitted with a
thermal bremsstrahlung of $kT\sim 0.92\pm 0.48$ keV plus a power-law
model of $\Gamma \sim 3.02\pm 0.14$, reduced $\chi^2=1.251$ (22
$d.o.f.$).  The derived hard X-ray non-thermal emission spectral
property is still consistent with the result by the Suzaku
observations from 13 -- 28 keV (Tamagawa et al. 2009).  The derived
continuum flux from 3 -- 100 keV is about 8.5$\times 10^{-11}$ erg
cm$^{-2}$ s$^{-1}$, corresponding to a hard X-ray luminosity of
$\sim 2\times 10^{35}d_4^2$ erg s$^{-1}$.

However, there is still some excess around 60-90 keV in the residuals from this continuum best-fit
model, which might be attributed to the $^{44}$Ti line emission. Thus we re-fit the spectra from 3 -- 200 keV with a
continuum model (a thermal bremsstrahlung and a power law) plus two
gaussian lines. The line positions in the fitting is fixed to be at
67.9 and 78.4 keV, and the line width is set to be zero due to the
low spectral resolution of IBIS/ISGRI around 70 keV ($\sim 8\%$). In
addition, to improve the statistical significance of hard X-ray line
detection, we also fix the line flux ratio during the fitting:
$F_{68}=0.93F_{78}$. Then we derive $kT\sim 0.81\pm 0.45$ keV, and
the photon index of $\Gamma\sim 3.01\pm 0.16$ with the mean line
flux of $F_{78}\sim (1.3\pm 0.5)\times 10^{-5}$ ph cm$^{-2}$
s$^{-1}$ (reduced $\chi^2=0.422~/21\ d.o.f.$). The significance of
the detection is still low ($\sim 2.6\sigma$) with the present measurements, so that we also
give a $3\sigma$ upper limit of $1.5\times 10^{-5}$ ph
cm$^{-2}$ s$^{-1}$ on the $^{44}$Ti line emission in Tycho. Anyway, this marginal detection of the $^{44}$Ti signal in
this Type Ia supernova remnant will be interesting and help us to
probe the progenitor of this remnant.

\begin{figure}
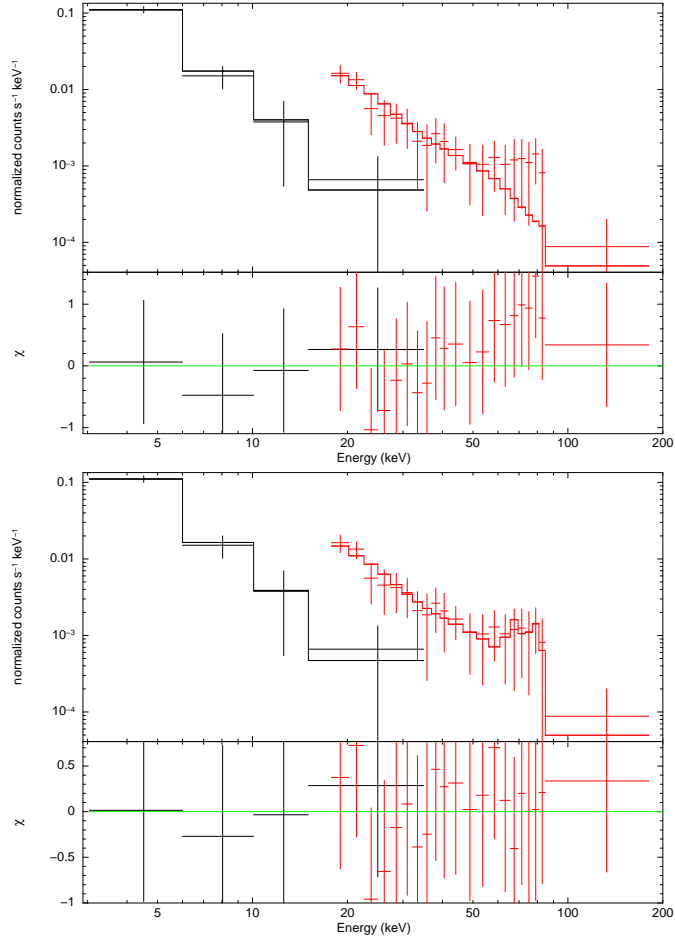

\centering
\includegraphics[angle=-90,width=9cm]{tycho_bp.eps}
\includegraphics[angle=-90,width=9cm]{tycho_bp_cyc.eps}
\caption{{\bf Top:} The spectrum of the Tycho SNR in hard X-ray band
of 3 -- 200 keV. The spectrum is fitted with a thermal
bremsstrahlung of $kT\sim 0.92\pm 0.48$ keV plus a power-law model
of $\Gamma \sim 3.02\pm 0.14$, reduced $\chi^2=1.251$ (22 $d.o.f.$).
The continuum flux of 3 -- 100 keV is $\sim (8.5\pm 0.5)\times
10^{-11}$ erg cm$^{-2}$ s$^{-1}$. {\bf Bottom:} The spectrum of the
Tycho SNR with two $^{44}$Ti emission lines at $\sim 68$ and 78 keV
fitted. See the text for details. }
\end{figure}

\section{Discussion}

\subsection{$^{44}$Ti amount in Tycho remnant}

\begin{figure}
\centering
\includegraphics[angle=0,width=12cm]{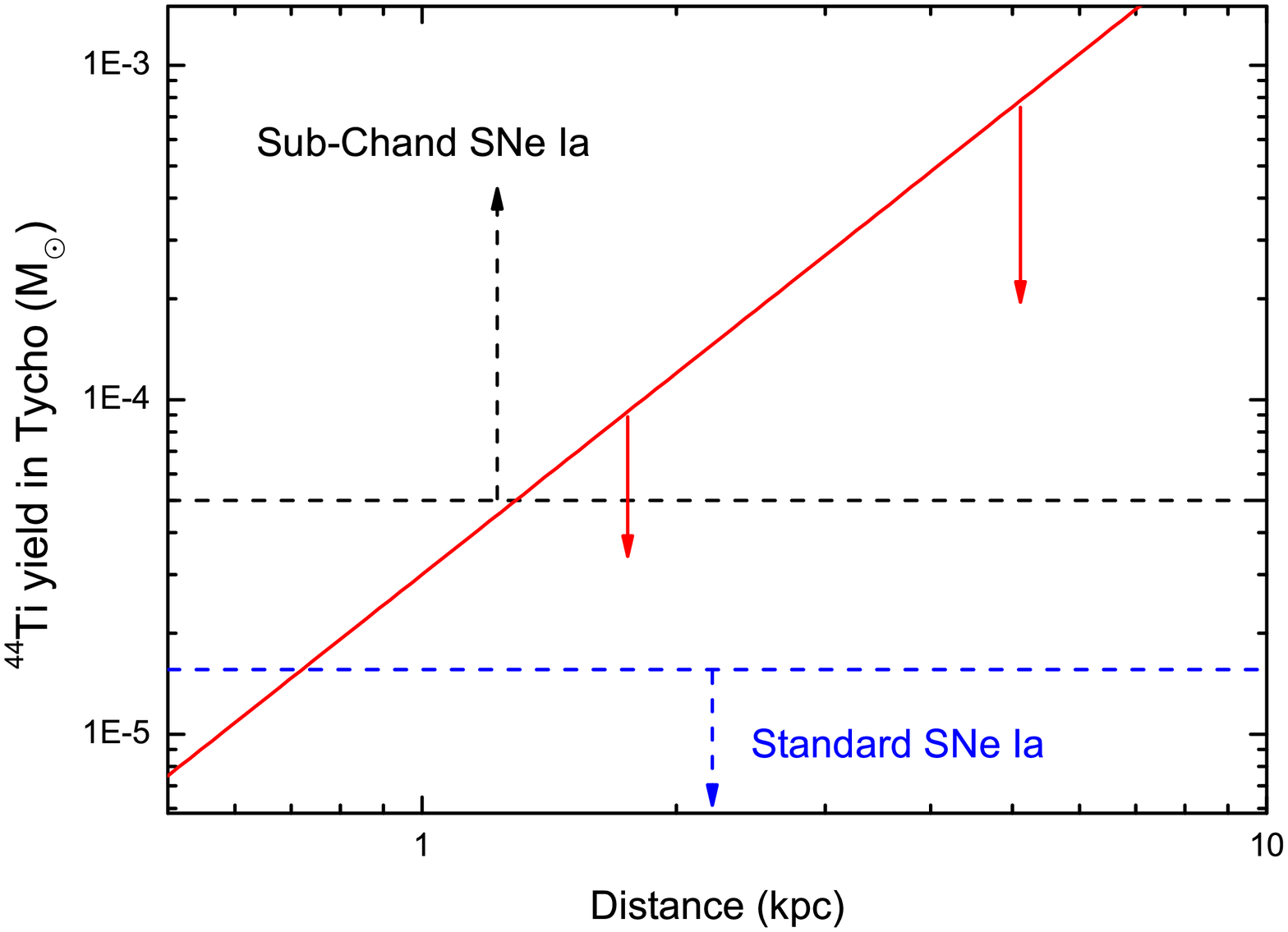}
\caption{The derived $^{44}$Ti yield upper limit in Tycho as function of
distance. The red solid line represents the upper
limit function (a $3\sigma$ upper limit of $1.5\times 10^{-5}$
ph cm$^{-2}$ s$^{-1}$) of the $^{44}$Ti yield. The figure also
displays the calculated $^{44}$Ti yield ranges from the 2D and 3D
simulation results of standard Chandrasekhar mass (about 1.4\ms,
Maeda et al. 2010; Seitenzahl et al. 2013) and sub-Chandrasekhar
mass models (Woosley \& Kasen 2011) of Type Ia supernovae.  }
\end{figure}

The creation of $^{44}$Ti in a supernova requires the presence of a
large mass fraction of helium heated briefly to temperatures in
excess of about 2 billion K. These conditions exist either when
material in nuclear statistical equilibrium is rapidly quenched at
such a low density that the helium fails to fully reassemble into
iron-group elements, the so called ``alpha-rich freeze out''
(Woosley et al. 1973), or when a detonation wave passes through a
helium-rich composition at typical white dwarf densities. The
alpha-rich freeze out can happen either in a massive star, where it
occurs in the deepest layers to be ejected (Timmes et al. 1996), or
in the carbon-rich layers of a Type Ia supernova (Iwamoto et al.
1999; Maeda et al. 2010; Seitenzahl et al. 2013). In the latter
case, the synthesis of $^{44}$Ti is relatively small due to the high
density during the freeze-out. Helium detonation occurs only in Type
Ia supernovae and there the production of $^{44}$Ti can sometimes be
very substantial (Woosley \& Weaver 1994; Timmes et al. 1996).

These general considerations translate into typical $^{44}$Ti yields
for current models for Type Ia supernovae. For Type Ia supernovae
resulting from carbon deflagration and detonation in white dwarfs
near the Chandrasekhar mass, the yield is quite low, typically $(0.2
- 1.6) \times 10^{-5}$ \ms based on the multi-dimensional
simulations (Maeda et al. 2010; Seitenzahl et al. 2013) with the
lower values more typical of the most recent three-dimensional (3D)
models. Sub-Chandrasekhar mass models for Type Ia supernovae, on the
other hand, are prolific sources of $^{44}$Ti and it has long been
thought that the production of the nucleus $^{44}$Ca in nature
occurs chiefly in this kind of explosion (Timmes et al. 1996). These
models are characterized by a shell of helium of about $0.05 - 0.2$
\ms atop a carbon-oxygen dwarf of 0.7 to 1.0 \ms. The detonation of
the helium induces a secondary detonation of the carbon and the
entire white dwarf explodes leaving no remnant. Recent calculations
(Woosley \& Kasen 2011) of this sort of model give $^{44}$Ti yields
in the range $(5 - 500)\times 10^{-5}$ \ms. While we can find no
published nucleosynthesis studies, we also expect that merging white
dwarfs in which one of the components is a helium white dwarf would
give similarly high yields provided that some portion of the helium
detonates (Dan et al. 2012).

Using the observed line flux, the $^{44}$Ti yield
synthesized in the supernova can be estimated: \beq M_{44Ti}\approx
4\pi d^2 44m_p \tau \exp(t/\tau) F_{44Ti}, \enq where $d$ is the
distance of the Tycho remnant, $m_p$, the proton mass, $\tau$, the
characteristic time of the $^{44}$Ti decay chain, $t$, the time
since the explosion, and $F_{44Ti}$, the flux of the $^{44}$Ti
emission line.  As discussed in the introduction, we take
the most likely distance distribution of the Tycho supernova remnant
to be 1.7 -- 5.0 kpc in this work.

An upper limit of the $^{44}$Ti line flux in Tycho is derived to be
$1.5\times 10^{-5}$ ph cm$^{-2}$ s$^{-1}$ ($3\sigma$). Fig. 4
shows the $^{44}$Ti yield upper limit (the solid line) we have
observed in Tycho remnant plotted against the uncertain distance to
the remnant according to Eq. (1). The region between two red arrows
is the estimated $^{44}$Ti yield limit for distances in the most
probable range, 1.7 -- 5.0 kpc, based on the measured $^{44}$Ti line
flux upper limit at 68 and 78 keV in Tycho.  The estimated $^{44}$Ti
yield ranges according to the simulation results of both the
standard Chandrasekhar mass models and sub-Chandrasekhar mass models
are also plotted. The present observed upper limit of the $^{44}$Ti
yield in Tycho is still consistent with both two explosion models
for the Type Ia supernovae.

\subsection{Non-thermal hard X-ray emission up to 100 keV}
Based on the diffusive shock acceleration (DSA) theory we discuss
the acceleration ability of Tycho SNR shock, using the hard X-ray
observation up to 100 keV, that probing the cutoff of the electron
spectrum.

The acceleration of electrons suffers from radiative energy loss.
The maximum synchrotron photon energy where the electron
acceleration and synchrotron cooling times are equal is $h\nu_{\rm
cutoff}\sim0.15\xi(E_{e,\max})^{-1}v_8^2\rm keV$ (e.g., Katz \&
Waxman 2008), where $\xi(E)\ge1$ is the ratio of the diffusion
coefficient to the Bohm diffusion one and could be energy dependent,
and $v_8$ is the SNR shock velocity in units of $10^8$ cm s$^{-1}$. The measurements of proper motion
of ejecta in Tycho SNR usually give an expansion velocity of
$v\sim5000$ km/s (Hayato et al. 2008; Katsuda et al. 2010), so the
synchrotron cutoff is $\sim4\xi^{-1}$keV, while the detected 100 keV
emission in Tycho is well above. Recently Zirakashvili \& Aharonian
(2007; 2010) investigated the spectral shape of the
shock-accelerated electrons subject to synchrotron cooling in the
context of DSA theory. They provided useful approximations for the
subsequent synchrotron spectral shape, which is a slow function
other than sharp cutoff. Using their approximation (eq. 37 in
Zirakashvili \& Aharonian 2007), in order for the 3--100 keV
emission to be statistically compatible to a power law with photon
index of $\sim3$, the cutoff should be $\sim3$~keV\footnote{The
cutoff energy in eq. (37) of Zirakashvili \& Aharonian (2007) is not
exactly the same as $h\nu_{\rm cutoff}$ here but different by about
$20\%$. However given the uncertainty in the DSA theory and for
purpose of order of magnitude estimate we neglect the difference.}.
Therefore the highest energy electrons are accelerated close to the
Bohm limit, $\xi(E_{e,\max})\sim1.3(v/{\rm 5000km\,
s^{-1}})^2(h\nu_{\rm cutoff}/3{\rm keV})^{-1}$, consistent with the
fact that the approximation used is derived for the Bohm diffusion
regime.

The postshock magnetic field can be constrained with the multi-band
synchrotron spectrum. If the accelerated electron distribution
follows a single power law, the downstream electron distribution is
a broken power law, with a cooling break in the synchrotron spectrum
corresponding to where the synchrotron cooling time is equal to the
SNR age, $h\nu_{\rm cool}\approx3B_{-4}^{-3}t_{\rm kyr}^{-2}$eV
(e.g., Katz \& Waxman 2008). At this break the synchrotron spectral
index steepens by a half. Given the normalization and spectral index
of radio emission from Tycho (Kothes et al. 2006), the cooling break
occurs around $\sim10$eV in order for the extrapolated flux to match
the observed X-ray flux of Suzaku and INTEGRAL (see also Fig. 10 in
Morlino \& Caprioli 2012). Given $h\nu_{\rm cool}\sim10$~eV and
$t\approx440$~yr, the postshock magnetic field is then derived to be
$B\sim120(h\nu_{\rm cool}/10{\rm eV})^{-1/3}\mu\rm G$, insensitive
to $\nu_{\rm cool}$ and comparable to that derived from the
observation of the X-ray rim (e.g., Warren et al.
2005)\footnote{Note that by the similar argument Morlino \& Caprioli
(2012) derived an even somewhat higher magnetic field of $200\mu G$,
leading to higher cosmic ray energy $E_{\max}$.}.

The acceleration of nuclei suffers mainly from the limited SNR age,
thus the maximum nuclei energy is
$E_{\max}\approx60Z\xi(E_{\max})^{-1}B_{-4}v_8^2t_{\rm kyr}$~TeV
(e.g., Katz \& Waxman 2008), where $Z$ is the nuclear charge number.
Using the above constraints of $B$ and $\xi(E_{e,\max})$ by
observations, the maximum cosmic ray energy can be expressed as
\begin{equation}
 E_{\max}\approx640Z\frac{\xi(E_{e,\max})}{\xi(E_{\max})}
 \left(\frac {h\nu_{\rm cutoff}}{3{\rm keV}}\right)\left(\frac{h\nu_{\rm cool}}{10{\rm eV}}\right)^{-1/3}
 \rm TeV,
\end{equation}
only with the unknown parameter of $\xi(E_{e,\max})/\xi(E_{\max})$.
With the constraint of $B$ the 100~keV emitting electrons have high
energy of $E>200$~TeV, not far from the above value. The fact that
the 3--100 keV spectrum is compatible to the case of Bohm diffusion
regime suggests $\xi(200\rm TeV)\sim1$. Since $\xi$ may not vary
sensitively with particle energy, we have $\xi(E_{\max})\sim1$ and
$\xi(E_{e,\max})/\xi(E_{\max})\sim1$, although
$\xi(E_{e,\max})/\xi(E_{\max})\ll1$ could not be ruled out. There
are hints from measurements of the expansion rate that Tycho is
currently transiting into the Sedov-Taylor phase (e.g., Katsuda et
al. 2010). During the transition SNRs are expected to produce the
highest energy cosmic rays in their whole lives. In summary, Tycho
may not accelerate protons up to the PeV scale; however, it is
possible that light nuclei with $Z\ga$ a few may be accelerated to
the PeV scale, but impossible to be far above PeV. Similar
conclusion has been reached by e.g., Bell (2013) in a specific
model.

\section{Conclusion}

This work studied the hard X-ray properties of the Tycho SNR using
INTEGRAL deep observations from 2003 -- 2011. We detected Tycho
firstly up to 90 keV. The X-ray spectrum from 3 -- 100 keV can be
described by a thermal bremsstrahlung of $kT\sim 0.8$ keV and a
power-law model of $\Gamma\sim 3$. A bump feature around 60 -- 90
keV is found in the spectrum which is possibly the signal of $^{44}$Ti emission lines at 68 and 78 keV. The gaussian line profile is
used to fit the two lines, and we find the marginal detection of the
$^{44}$Ti lines at a significance level of $\sim 2.6\sigma$. Thus we
find a $3\sigma$ upper limit of $1.5\times 10^{-5}$ ph cm$^{-2}$
s$^{-1}$ for the $^{44}$Ti lines.

The detected non-thermal emission up to $\sim 90$ keV in the Tycho
SNR also suggests that the remnant could accelerate protons to at
least $\sim 200$ TeV, but not up to the PeV scale. The light nuclei
with $Z\ga$ a few may be accelerated to the PeV scale, around the
cosmic ray spectral ``knee" region. This implies that the
composition of cosmic rays at the ``knee" is not dominated by
protons but by light nuclei, if normal SNRs are the origins of
cosmic rays at the ``knee".

\begin{acknowledgements}
We are grateful to the referee for the critical and fruitful
comments and Stan Woosley for discussions on nucleosynthesis in
supernovae. This work is supported by the NSFC (No. 11073030,
11273005), the SRFDP (20120001110064), the CAS Open Research Program
of Key Laboratory for the Structure and Evolution of Celestial
Objects, and the National Basic Research Program (973 Program) of
China (2014CB845800).

\end{acknowledgements}

\end{document}